\documentclass[twocolumn,prx,preprintnumbers,superscriptaddress,amsmath,amssymb,english]{revtex4}
\usepackage{amsmath,amssymb,amsfonts,mathrsfs}
\usepackage{amsthm}
\usepackage{CJK}
\usepackage{bm}
\usepackage{babel}
\usepackage{graphicx}
\allowdisplaybreaks
\usepackage{braket}
\usepackage[breaklinks]{hyperref}
\usepackage{color}
\usepackage{float}
\hypersetup{colorlinks=true, linkcolor=blue, citecolor=blue, filecolor=blue, urlcolor=blue}
\UseRawInputEncoding

\begin{document}

\title{Symmetry-protected Dirac nodal lines and large spin Hall effect in $\mathbf{V_6Sb_4}$ with kagome bilayer}

\author{Y. Yang}
\affiliation{CAS Key Laboratory of Strongly-coupled Quantum Matter Physics, Department of Physics, University of Science and Technology of China, Hefei, Anhui 230026, China}

\author{R. Wang}
\email[]{rcwang@cqu.edu.cn}
\affiliation{Institute for Structure and Function $\&$ Department of physics, Chongqing University, Chongqing 400044, China}
\affiliation{Center of Quantum Materials and Devices, Chongqing University, Chongqing 400044, China}
\affiliation{Chongqing Key Laboratory for Strongly Coupled Physics,  Chongqing University, Chongqing 400044, China}

\author{M.-Z. Shi}
\affiliation{CAS Key Laboratory of Strongly-coupled Quantum Matter Physics, Department of Physics, University of Science and Technology of China, Hefei, Anhui 230026, China}

\author{Z. Wang}
\affiliation{CAS Key Laboratory of Strongly-coupled Quantum Matter Physics, Department of Physics, University of Science and Technology of China, Hefei, Anhui 230026, China}
\affiliation{CAS Center for Excellence in Superconducting Electronics (CENSE), Shanghai 200050, China}

\author{Z. Xiang}
\affiliation{CAS Key Laboratory of Strongly-coupled Quantum Matter Physics, Department of Physics, University of Science and Technology of China, Hefei, Anhui 230026, China}

\author{X.-H. Chen}
\affiliation{CAS Key Laboratory of Strongly-coupled Quantum Matter Physics, Department of Physics, University of Science and Technology of China, Hefei, Anhui 230026, China}
\affiliation{CAS Center for Excellence in Superconducting Electronics (CENSE), Shanghai 200050, China}
\affiliation{CAS Center for Excellence in Quantum Information and Quantum Physics, Hefei, Anhui 230026, China}
\affiliation{Collaborative Innovation Center of Advanced Microstructures, Nanjing University, Nanjing 210093, China}

\begin{abstract}
Recently, a family of nonmagnetic kagome metals \textit{A}$\mathrm{V_3Sb_5}$ (\textit{A}=K, Rb, and Cs) has attracted significant attention for realizing the intertwining of quantum order and nontrivial topology. However, these compounds have been identified to host complex band structures. Therefore, it is desirable to design and synthesize novel kagome materials with a simple band topology and good transport properties. In this study, using first-principles calculations, we present the electronic properties and the intrinsic spin Hall effect  of V$_6$Sb$_4$, the latest experimentally synthesized vanadium-based compounds with a kagome bilayer. In the absence of spin-orbital coupling (SOC), this compound is a Dirac nodal line semimetal with symmetry-protected nodal rings near the Fermi level. Within the SOC, the spin-rotation symmetry breaks the gaps of the nodal rings with a small band gap. Furthermore, based on the Wannier tight-binding approach and the Kubo formula, we propose a large spin Hall effect in V$_6$Sb$_4$, which intrinsically originates from the spin Berry curvature. Our work further expands nonmagnetic kagome compounds for applications in spintronics accompanied by exotic quantum order.
\end{abstract}



\maketitle
\section{Introduction}
Materials with a kagome lattice have attracted significant interest in condensed matter physics owing to their unusual lattice geometry. The kagome lattice possesses a two-dimensional (2D) corner-sharing triangular network, which ensures that its band structure uniquely exhibits the coexistence of linear Dirac and flat bands. Consequently, numerous topological and many-body quantum phenomena have been identified over the past decade, such as charge density waves (CDWs) \cite{PhysRevB.80.113102}, bond density wave order \cite{PhysRevB.81.235115,PhysRevB.90.035118}, charge fractionalization \cite{PhysRevB.83.165118}, quantum limit-Chern topological phases \cite{2020Quantum}, quantum spin-liquid states \cite{RevModPhys.89.025003,spin-liquid}, and topological superconductivity \cite{PhysRevB.79.214502}. These findings strengthen the understanding of correlated band topology.

Recently, the discovery of a new family of nonmagnetic kagome metals \textit{A}$\mathrm{V_3Sb_5}$ (\textit{A}=K, Rb, and Cs) further promoted the development of exotic physics in kagome lattices \cite{PhysRevMaterials.3.094407, NatuecomChen, PhysRevX.11.031050, PhysRevX.11.031026, PhysRevLett.127.046401, PhysRevLett.125.247002, PhysRevMaterials.5.034801}. These compounds have a superconducting ground state intertwined with the CDW order \cite{PhysRevX.11.031050, PhysRevX.11.031026, PhysRevLett.127.046401}. Theoretical calculations have shown that the normal state of these compounds has been categorized as a $\mathbb{Z}_2$ topological metal. More strikingly, intense studies have been conducted to reveal the close correlations between superconductivity, CDW, and nontrivial band topology in \textit{A}$\mathrm{V_3Sb_5}$ \cite{PhysRevLett.125.247002, PhysRevMaterials.5.034801,2021chiral,2021Naturegao, 2021Natureze}, pushing this topic to the quantum frontier. Beyond the acquisition of promising quantum phenomena in \textit{A}$\mathrm{V_3Sb_5}$, of equal importance is the exploration of more kagome materials, which will further lead to the experimental realization of rich phenomena between band topology and quantum ordered states. Additionally, the complicated Fermi surface topology of \textit{A}$\mathrm{V_3Sb_5}$, such as multiple trivial and nontrivial bands across the Fermi level,  would make the understanding of transport properties ambiguous. For instance, a giant anomalous Hall effect in absence of a magnetic order was observed at low temperatures \cite{PhysRevB.104.L041103,sciadvabb6003}, but the mechanism remained unclear \cite{PhysRevB.103.L241117, FENG20211384}. Therefore, the design and synthesis of novel kagome materials with ideal band topologies and good transport properties are desirable.

To address this issue, we first reviewed the crystal structure of \textit{A}$\mathrm{V_3Sb_5}$. The \textit{A}$\mathrm{V_3Sb_5}$ compounds crystallized in the space group P6/mmm (No. 191) \cite{PhysRevMaterials.3.094407}, possessing a layered structure with a prototypical kagome V$_3$Sb layer. This layer is composed of a V-kagome sublattice interpenetrated with the Sb ( Sb1) trigonal sublattice. The other Sb ( Sb2) sublattice encapsulates the V$_3$Sb kagome layer;  meanwhile, the alkali \textit{A} layer only fills the natural space left between the Sb2 layers, thus it does not contribute to the bands near the Fermi level. 
Consequently, the electronic and topological properties of \textit{A}$\mathrm{V_3Sb_5}$ are dominated by the kagome V$_3$Sb monolayer. However, to the best of our knowledge, vanadium-based materials with kagome bilayer or beyond have rarely been reported in the literature.

Recently, our group discovered two members of vanadium compounds that contain a $(\mathrm{V_3Sb})_2$ kagome bilayer in their crystal structures with the space group $R\bar{3}m$ (No. 166), such as \textit{A}$\mathrm{V_6Sb_6}$ and V$_6$Sb$_4$ \cite{A166arXiv}. Importantly these two novel members and \textit{A}$\mathrm{V_3Sb_5}$, can be represented as a generic chemical formula \textit{A}$_{m-1}$Sb$_{2m}$(V$_3$Sb)$_n$ (i.e., \textit{A}$\mathrm{V_3Sb_5}$ with $m=2$, $n=1$ , \textit{A}$\mathrm{V_6Sb_6}$ with $m=2$, $n=2$, $\mathrm{V_6Sb_4}$ with $m=1$, $n=2$). Importantly, we demonstrated that \textit{A}$\mathrm{V_6Sb_6}$ compounds host exotic type-II nodal lines adjacent to the Fermi level and superconductivity under pressure \cite{A166arXiv, PhysRevB.104.245128}, implying that vanadium-based kagome bilayer compounds could potentially be viewed as ideal candidates for studying exotic quantum phenomena with unconventional fermionic excitations.

In this work, using first-principles calculations, we investigated the band topology and its induced intrinsic spin Hall effect (SHE) of V$_6$Sb$_4$. In the absence of spin-orbital coupling (SOC), compound V$_6$Sb$_4$ is also a symmetry-protected Dirac nodal line semimetal. 
Distinguished from the open type-II nodal lines threading the Brillouin zone (BZ) in \textit{A}$\mathrm{V_6Sb_6}$, V$_6$Sb$_4$ possesses closed type-I Dirac nodal rings with respect to the inverted point at the boundary of the BZ. When SOC is considered, the breaking of spin-rotation symmetry destroys the nodal rings. Strikingly, the spin Berry curvature of gapped nodal rings can lead to a large intrinsic SHE, which indicates that the V$_6$Sb$_4$ compound can create a transverse pure spin current under a longitudinal electric field \cite{RevModPhys.87.1213}. Therefore, our results further expand the nonmagnetic kagome compounds for applications in spintronics accompanied by exotic quantum order.

\begin{figure}
    \centering
    \includegraphics[scale=0.5]{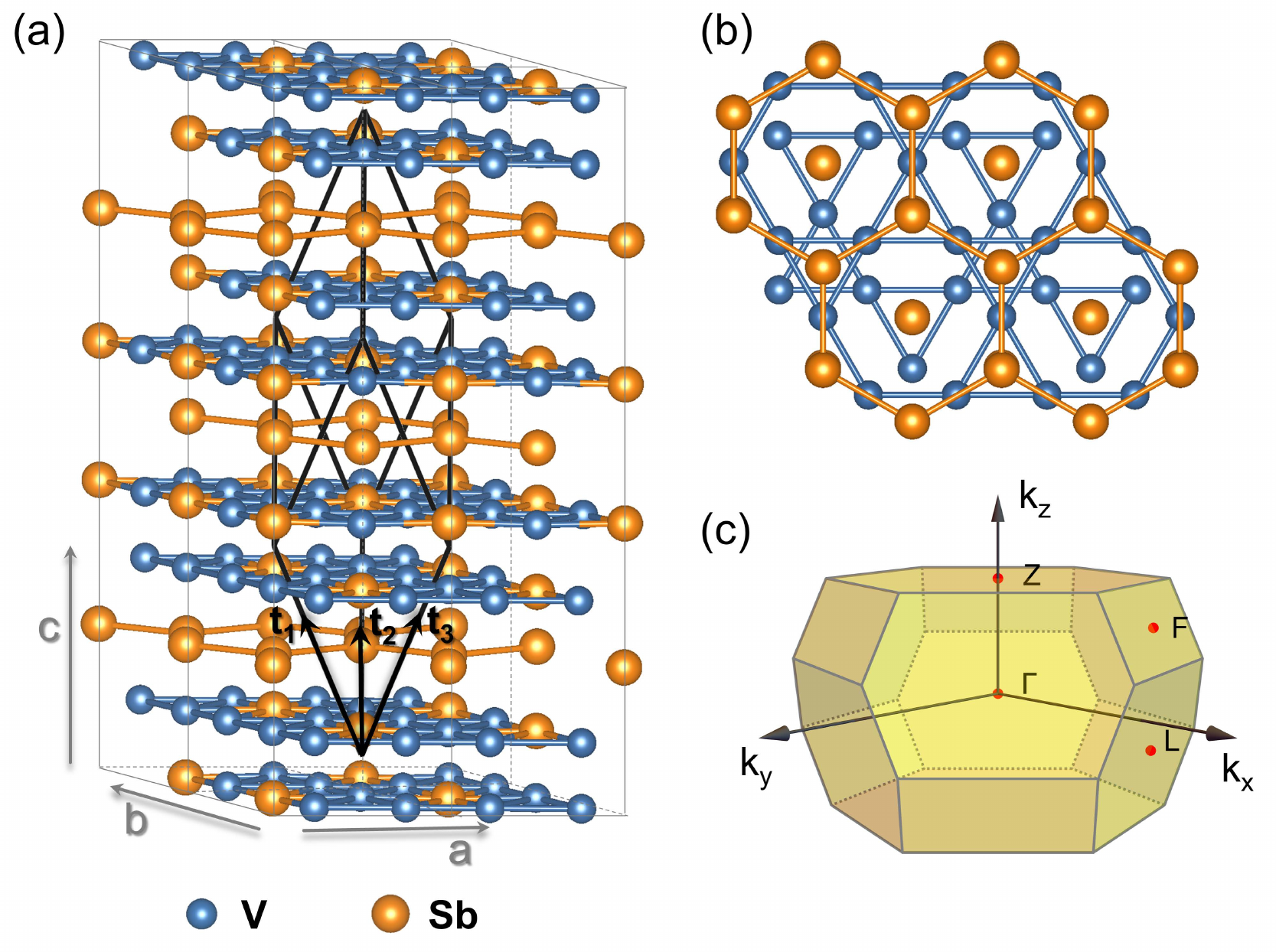}
    \caption{(a) Crystal structure of V$_6$Sb$_4$ with a vanadium-based kagome bilayer. The conventional unit cell (gray lines) in the hexagonal representation and primitive unit cell (black lines) in the rhombohedral representation are illustrated. (b) Top view of V$_6$Sb$_4$ compound. It was found that two $\mathrm{V_3Sb}$ bilayers were separated by two Sb (i.e., Sb2) layers. The V atoms in each $\mathrm{V_3Sb}$ bilayer form a kagome lattice, and the Sb (i.e., Sb1) atoms in each layer form a honeycomb lattice. (c) Bulk Brillouin zone (BZ) of a rhombohedral unit cell in which the high-symmetry points are marked.
    \label{FIG1}}
\end{figure}

\section{Computational Methods}

\begin{figure*}
    \centering
    \includegraphics[scale=0.5]{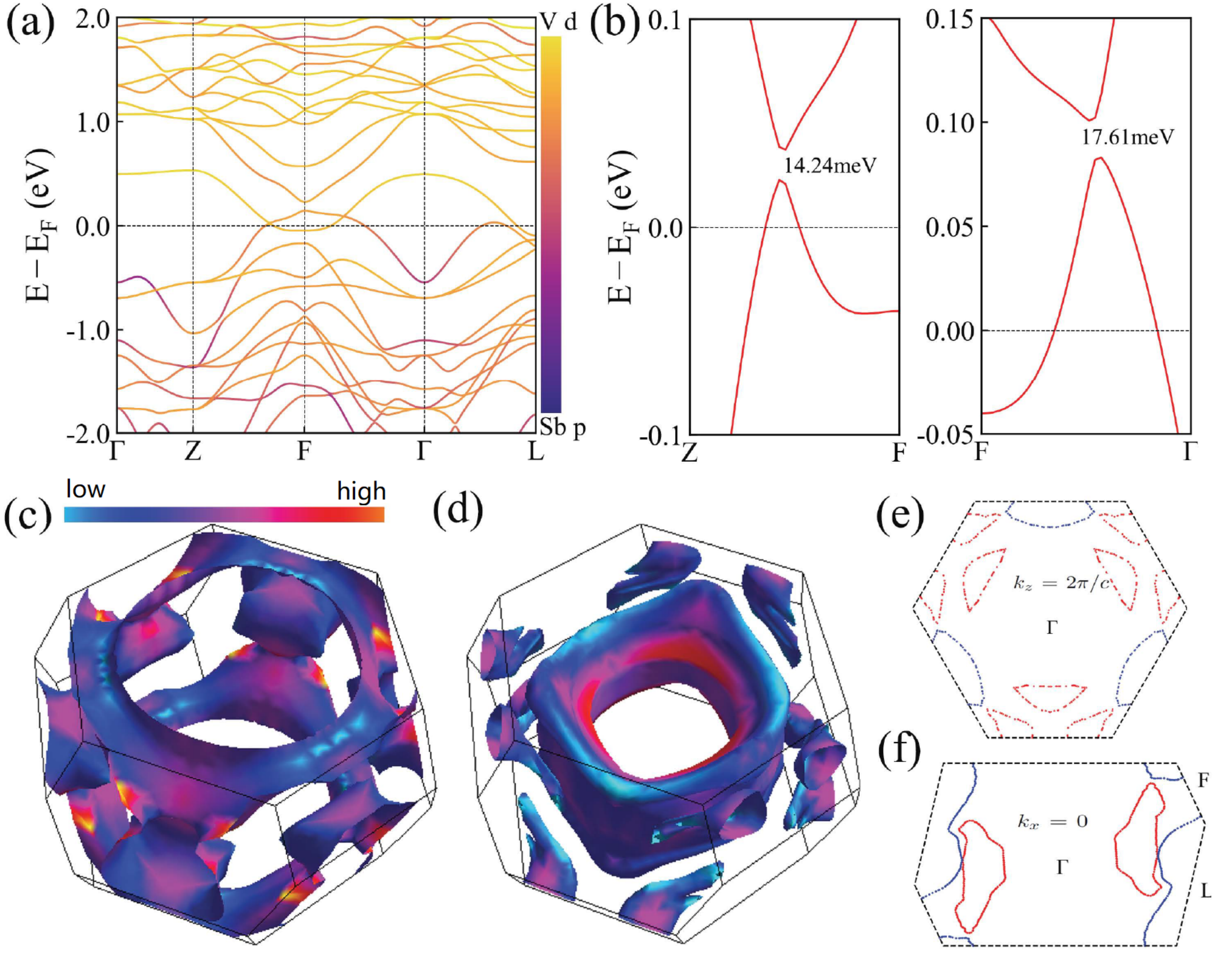}
    \caption{Electronic band structures and Fermi surface of V$_6$Sb$_4$. (a) Orbital resolved electronic band structures along high-symmetry paths in the absence of SOC. The yellow and blue colors represent the components of the V $d$ and Sb $p$ orbitals, respectively. (b) Enlarged views of band structures in the $F$-$Z$ and $F$-$\Gamma$ directions in presence of SOC. The three-dimensional Fermi surface of (c) electron and (d) hole pockets. The color bar represents Fermi velocity. Two-dimensional slices of the Fermi surface at (e) $k_z=2\pi/c$ and (f) $k_x=0$. The red and blue regions denote the hole and electron pockets, respectively.
    \label{FIG2}}
\end{figure*}

To investigate the electronic and topological properties of V$_6$Sb$_4$, we performed first-principles calculations as implemented in the Vienna \textit{ab initio} simulation package \cite{PhysRevB.54.11169} based on density functional theory (DFT) \cite{PhysRev.140.A1133}. We employed the generalized gradient approximation (GGA) with Perdew-Burke-Ernzerhof formalism to describe the exchange-correlation functional \cite{PhysRevLett.77.3865}. The valence electron-ion interaction was treated by projector-augmented-wave potentials \cite{PhysRevB.59.1758}. The cutoff energy of the plane wave basis was set to 500 eV, and the BZ was sampled using a $15\times15\times15$ Monkhorst-Pack grid \cite{PhysRevB.13.5188}. To reveal the van der Waals interactions along the $c$-layer stacking direction, the Crimme (DFT-D3) method was used \cite{D3}. The forces on each atom were relaxed to less than $1.0\times 10^{-4}$ eV/{\AA}. The topological classification was confirmed by the $\mathbb{Z}_2$ invariant \cite{PhysRevLett.98.106803}, which was calculated from the parity eigenvalues at the time-reversal invariant momenta (TRIM) points using the IRVSP package \cite{GAO2021107760}. By projecting the Bloch states into highly symmetric atomic orbitals of Wannier functions (WF) (Sb-$p$, V-$p$, and V-$d$ orbitals), we constructed a Wannier tight-binding (TB) Hamiltonian based on maximally localized Wannier function (MLWF) methods \cite{Mostofi2008,RevModPhys.84.1419}. For the construction of the MLWF, a large inner-frozen window of -8.0 to 8.0 eV with respect to the Fermi level was used to calculate the intrinsic spin Hall conductivity (SHC). Then, we calculated the SHC by employing the Kubo formula in the clean limit \cite{PhysRevLett.94.226601, PhysRevB.98.214402},
\begin{equation}\label{eqc}
\begin{split}
\sigma_{\alpha \beta}^{\gamma} & =e\hbar \int {\frac{dk_x dk_y dk_z}{(2\pi)^3}}\sum_{n}f_{n\mathbf{k}}\Omega^{\gamma}_{n,\alpha \beta}(\mathbf{k}), \\
\Omega^{\gamma}_{n,\alpha \beta}(\mathbf{k}) &= -2\text{Im} \sum_{m \neq n} \frac{ \big< \psi_{n\mathbf{k}} | {j}_{\alpha}^{\gamma} | \psi_{m\mathbf{k}} \big> \big< \psi_{m\mathbf{k}} | {v}_{\beta} | \psi_{n\mathbf{k}} \big>}{(E_{m\mathbf{k}}-E_{n\mathbf{k}})^{2}},
\end{split}
\end{equation}
where $f_{n\mathbf{k}}$ is the Fermi-Dirac distribution of the $n$th band, the spin current operator ${j}_{\alpha}^{\gamma}=1/2\{{v}_{\alpha},{s}^{\gamma}\}$ with the spin operator ${s}^{\gamma}$, and the velocity operator ${v}_{\alpha}=\frac{1}{\hbar}\frac{\partial H}{\partial k_{\alpha}}$ ($\alpha, \beta, \gamma = x, y, z$), and $E_{n\mathbf{k}}$ is the eigenvalue of the Bloch functions $\psi_{n\mathbf{k}}$. $\Omega^{\gamma}_{n,\alpha \beta}(\mathbf{k})$ denotes the spin Berry curvature. The SHC $\sigma_{\alpha \beta}^{\gamma}$ represents the spin current ${j}_{\alpha}^{\gamma}$ along the $\alpha$ direction with spin polarization along the $\gamma$ direction, which is generated by an electric field $E_{\beta}$ along the $\beta$ direction, that is, ${j}_{\alpha}^{\gamma}= \sigma_{\alpha \beta}^{\gamma} E_{\beta}$. From the Wannier TB Hamiltonian, we numerically calculated the integral of the spin Berry curvature in the momentum space to obtain the SHC based on Eq. (\ref{eqc}), and a dense $k$-mesh grid of $200 \times 200 \times 200$ was adopted.

\section{Results and Discussion}
\subsection{The crystal structure}
The V$_6$Sb$_4$ compound crystallizes in a rhombohedral layered structure with space group $R\bar{3}m$ (No. 166). As shown in Fig. \ref{FIG1}(a), its conventional unit cell in the hexagonal representation exhibits a layered structure with in-plane unite cell vectors $\mathbf{a}$ and $\mathbf{b}$, and $|\mathbf{a}|=|\mathbf{b}|=a$. The stacking period of V$_6$Sb$_4$ is characterized by the out-of-plane lattice vector $\mathbf{c}$. Therefore, the primitive lattice vectors of the unit cell in the rhombohedral representation link the hexagonal lattice constants as
\begin{equation}
\begin{split}
&\mathbf{t}_1=-\frac{a}{2}\mathbf{i} -\frac{\sqrt{3}}{6}a \mathbf{j}+ \frac{c}{3}\mathbf{k},\\
&\mathbf{t}_2=\frac{a}{2}\mathbf{i} -\frac{\sqrt{3}}{6}a \mathbf{j}+ \frac{c}{3}\mathbf{k},\\
&\mathbf{t}_3= \frac{\sqrt{3}}{3}a \mathbf{j}+ \frac{c}{3}\mathbf{k}.
\end{split}
\end{equation}
As shown in Figs. \ref{FIG1}(a) and \ref{FIG1}(b), the $(\mathrm{V_3Sb})_2$ bilayer forms kagome nets of V atoms coordinated by Sb1 atoms, and separated by the Sb2 sublattice with a honeycomb sublattice, forming a sandwiched structure. Each layer of the $(\mathrm{V_3Sb})_2$ bilayer is composed of a V-kagome sublattice interpenetrated with the Sb ( Sb1) trigonal sublattice. The V sublattice of each layer of the $(\mathrm{V_3Sb})_2$ bilayer forms a kagome lattice, which is analogous to \textit{A}$\mathrm{V_3Sb_5}$. The triangle sizes of the V kagome plane are divided into two unequal values induced by the breathing anisotropy, similar to Fe$_3$Sn$_2$ \cite{PhysRevLett.121.096401, NatureFeSn}. After full relaxation, the calculated lattice constants are $a = 5.42$ {\AA} and $c =20.15$ {\AA}, which are consistent with the experimental values \cite{A166arXiv}. The atom positions of the V$_6$Sb$_4$ compound are  Wyckoff $6h$ of the V atom, $2c$ of the Sb1 atom, and $2c$ of the Sb2 atom. The bulk BZ was shaped by a truncated octahedron, and the time-reversal invariant momentum (TRIM) points are denoted in the BZ [see Fig. \ref{FIG1}(c)].

\subsection{Electronic band structures and Fermi surface}

Based on first-principles calculations, we obtained the electronic band structures of the V$_6$Sb$_4$ compound and confirmed that nodal rings exist in the BZ. The band structures in absence of the SOC along the high-symmetry direction are shown in Fig. \ref{FIG2}(a). It was determined that band crossings in the $F$-$Z$ and $F$-$\Gamma$ directions at $\sim$0.03 meV and $\sim$0.09 meV above the Fermi level, respectively, formed hole pockets. The orbital-resolved contributions are also shown in Fig. \ref{FIG2}(a). The figure shows that the V $d$ orbitals dominate the bands around the Fermi level. Importantly, a visible band inversion occurred at the $F$ point, at which the occupied and unoccupied bands belong to the irreducible representations (IR) $A_u$ and $B_g$ of the $C_{2h}$ group. By further checking the small groups of the BZ, we determined that the two crossing bands belong to different IR $\Gamma_1$ and $\Gamma_2$ of $C_s$.  The mirror symmetry $C_s$ in the $k_{y}-k_{z}$ plane with $k_x=0$ (i.e., $M_x$) was derived from the product of the $I$-symmetry and two-fold rotational symmetry $\mathcal{C}_{2x}$ along the $x$ direction, that is, $M_x=I C_{2x}$. We further calculated the energy difference between the lowest conduction band and the highest valence band in the $k_{y}-k_{z}$ plane with $k_x=0$, and the calculated results indicate that the two bands inverted at the $F$ point formed continuous nodal rings (see Fig. S1 \cite{SM}). Consequently, nodal rings with respect to mirror-reflection symmetry could be present. Notably, these nodal rings of V$_6$Sb$_4$ were protected by the coexistence of spatial inversion ($I$) and time-reversal ($T$) symmetries, although they were located in the reflection-invariant plane. For the $IT$-protected nodal line semimetals, the spin-rotation symmetry enabled us to effectively treat electrons as spinless fermions
with $(IT)^2 = 1$ in absence of the SOC, corresponding to the $\mathbb{Z}_2$ classification for V$_6$Sb$_4$ \cite{PhysRevLett.116.156402}. Considering the three-fold rotational symmetry $C_{3z}$, there were three equivalent mirror planes, thus, six nodal rings enclosed the $F$ point in the BZ.

When the SOC effect was considered, the spin rotation symmetry was broken, which usually destroys nodal lines protected by the space-time $IT$ symmetry (i.e., $(IT)^2 = -1$) \cite{PhysRevLett.115.036806, PhysRevLett.116.156402}. As shown in Fig. \ref{FIG2}(b), the presence of $I$-symmetry forces the degeneracy of two spin bands within SOC, and the previous crossing bands belong to the same IR $\Gamma_4$ of $C_s$ now.  The DFT calculations confirmed that the band gaps are 14.2 meV along $F$-$Z$ and 17.6 meV along $F$-$\Gamma$. At point $F$, the occupied and unoccupied states near the Fermi level belong to the IRs of $\Gamma_3^- \oplus \Gamma_4^-$ and $\Gamma_3^+ \oplus \Gamma_4^+$ of $C_{2h}$, respectively. This opposite parity indicates that the band inverted feature at $F$  was preserved in presence of the SOC. In this case, we can use parity eigenvalues at the time-reversal invariant momenta (TRIM) points to depict the topological features between each pair of bands near the Fermi level. As illustrated in Table S1 \cite{SM}, the calculated parity eigenvalues indicate that bands near the Fermi level exhibit a topological switch, which was expected to generate exotic quantum transport properties.

Evidently, \textit{A}$\mathrm{V_3Sb_5}$ exhibits rich quantum oscillations \cite{PhysRevX.11.041030}, such as the multiple portions of Fermi surfaces that induce complex superposition of quantum oscillations.
Here, we plotted the three-dimensional (3D) Fermi surface of the electron and hole pockets in Figs. \ref{FIG2}(c) and \ref{FIG2}(d), respectively, where the color bar represents the Fermi velocity. The Fermi surface exhibits a strong 2D feature with respect to the three-fold rotational symmetry. The hole pockets of the Fermi surface exhibit a hollow-cylinder profile enclosed by the $\Gamma$ point, while the electron pockets are close to the boundary of the BZ. Near each $F$ point, clearly, there were two-hole pockets  related to the mirror symmetry, which originate from the crossing bands of nodal rings near the Fermi level. To further understand the properties of the Fermi surface, we provided 2D slices of the Fermi surface at $k_z=2\pi/c$ and $k_x=0$ in Figs. \ref{FIG2}(e) and \ref{FIG2}(f), respectively.

\subsection{Spin Hall effect}

\begin{figure}
    \centering
    \includegraphics[scale=0.29]{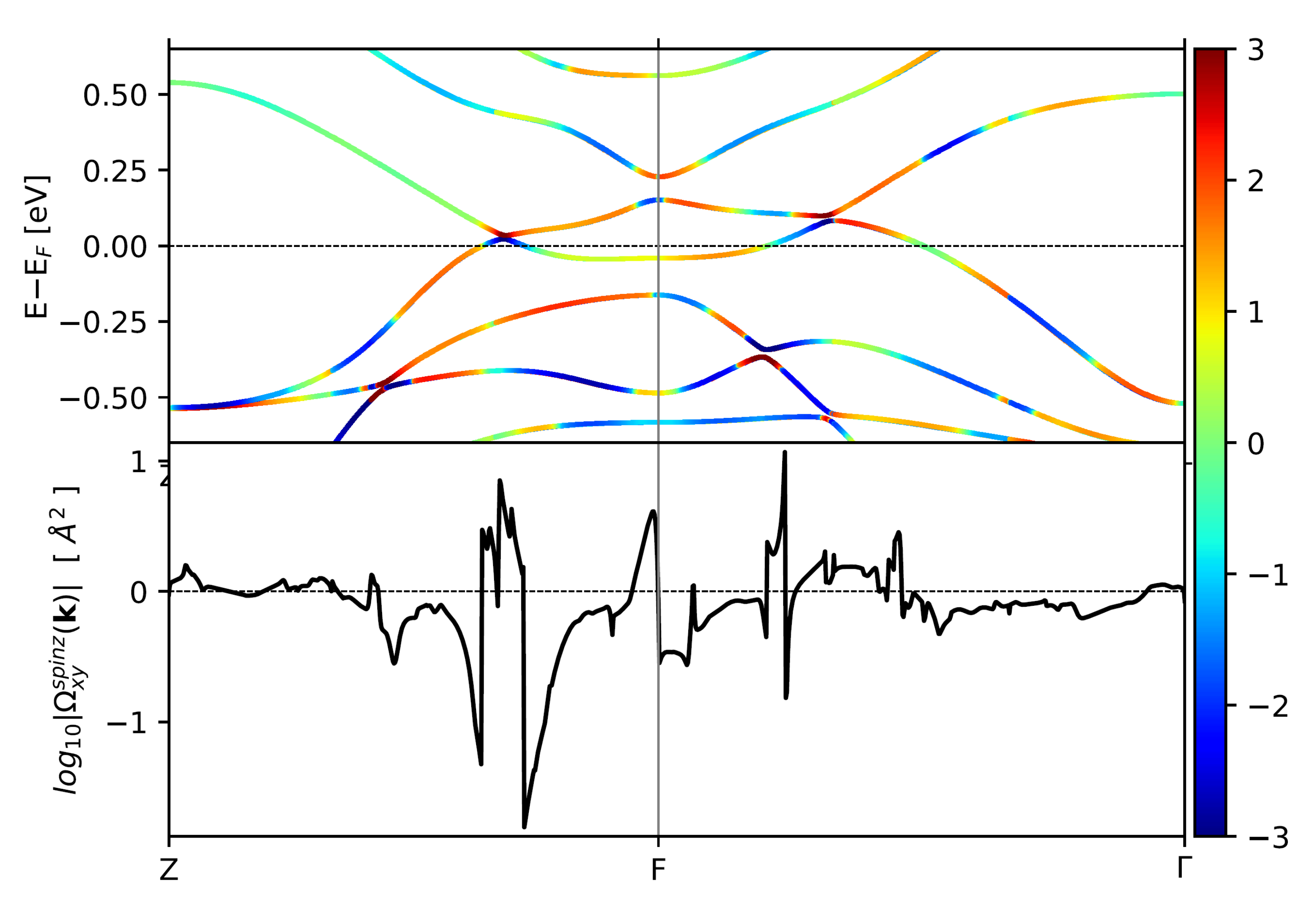}
    \caption{(a) Band-projected SHC of the V$_6$Sb$_4$ compound along the high-symmetry lines $Z$-$F$-$\Gamma$. The color bar in panel (a) is the weight of the SHC projected onto each band. The magnitude of the SHC projected on each band was taken using the logarithm \cite{PhysRevB.98.214402}. (b) Moment-resolved SHC along the high-symmetry lines $Z$-$F$-$\Gamma$.
    \label{FIG3}}
\end{figure}

Based on Eq. (\ref{eqc}), the intrinsic SHC is derived from the spin Berry curvature of $\Omega^{\gamma}_{n,\alpha \beta}(\mathbf{k})$ with the band index $n$ in presence of the SOC. Clearly, the band anticrossing induced by SOC can lead to large $\Omega^{\gamma}_{n,\alpha \beta}(\mathbf{k})$ \cite{PhysRevLett.117.146403}. To obtain a large SHC, an increase in the number of band anticrossing points (i.e., band crossings without SOC) is desirable. Therefore, an assembly of continual nodal points (i.e., nodal lines) in the BZ of V$_6$Sb$_4$ can be expected to induce a strong SHE. Further, the nontrivial bands near the Fermi level can also lead to a singularity in the momentum space, which may further enlarge the SHC of V$_6$Sb$_4$. To verify this, we calculated the band-projected plot of SHC along the high-symmetry lines $Z$-$F$-$\Gamma$, as shown in Fig. \ref{FIG3}(a). Here, the magnitude of the SHC projected on each band was taken by the logarithm \cite{PhysRevB.98.214402}. As expected, we found that the SHC was mostly concentrated near the region of the band gaps opened by SOC. As shown in Fig. \ref{FIG3}(b), we calculated the $k$-point resolved SHC, which can be expressed as
\begin{equation}
\Omega^{\mathrm{spin}z}_{x y}(\mathbf{k})=\sum_{n} f_{n\mathbf{k}}\Omega_{n,xy}^{\mathrm{spin}z}(\mathbf{k}).
\end{equation}
The figure shows that the peaks around the band gaps opened by SOC significantly contributed to SHC.

\begin{figure}
    \centering
    \includegraphics[scale=0.55]{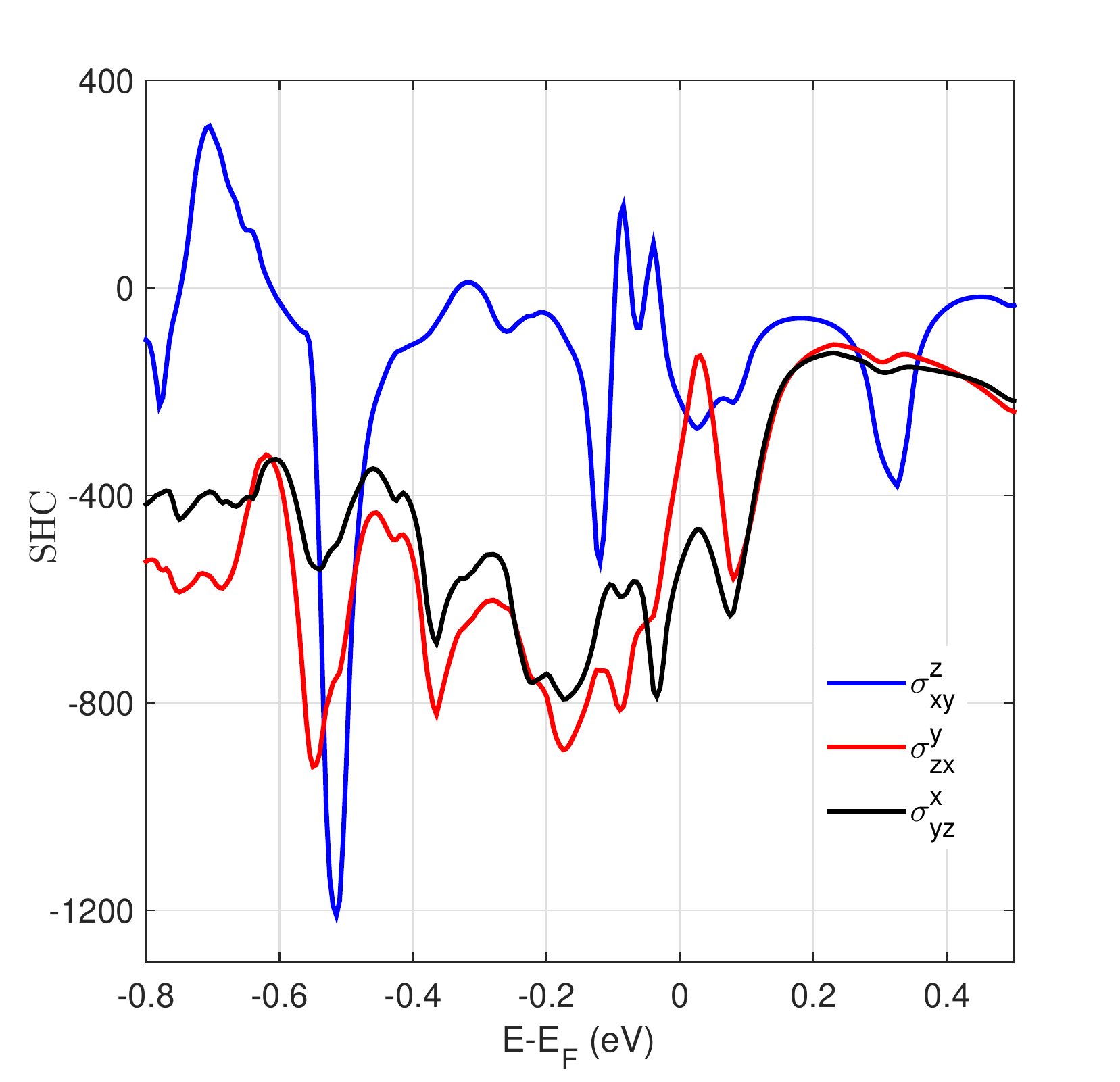}
    \caption{SHC of  V$_6$Sb$_4$ as a function of chemical potential. The independent SHC tensor elements $\sigma_{x y}^{z}$,  $\sigma_{zx}^{y}$, and $\sigma_{yz}^{x}$ are colored in blue, red, and black, respectively.
    \label{FIG4}}
\end{figure}

After analyzing the mechanism by the band-resolved and momentum-resolved SHCs, we investigated the intrinsic SHC of V$_6$Sb$_4$. The SHC $\sigma_{\alpha \beta}^{\gamma}$ is a third-order tensor with 27 elements.  The crystal structure of V$_6$Sb$_4$ belongs to the hexagonal lattice (space group $R\bar{3}m$, No. 166). Thus, SHE is anisotropic. The nonzero elements of the SHC depend on the space group of V$_6$Sb$_4$. Constrained by the crystal symmetry and time-reversal symmetry, there were only three independent nonzero elements for the V$_6$Sb$_4$ compound, that is,  $\sigma_{x y}^{z}$,  $\sigma_{zx}^{y}$, and $\sigma_{yz}^{x}$, which satisfy $\sigma_{x y}^{z}=-\sigma_{yx}^{z}$, $\sigma_{zx}^{y}=-\sigma_{zy}^{x}$, and $\sigma_{yz}^{x}=-\sigma_{xz}^{y}$,respectively. The SHC of  V$_6$Sb$_4$ as a function of the chemical potential is shown in Fig. \ref{FIG4}. Clearly, $\sigma_{yz}^{x}$ has the largest value of approximately 537 $(\hbar/e)(\Omega\cdot \mathrm{cm})^{-1}$ at the Fermi level, which is consistent with the nodal rings lying in the $k_y$-$k_z$ plane with $k_z=0$. In comparison, $\sigma_{x y}^{z}$ and $\sigma_{zx}^{y}$ have values of approximately 204 $(\hbar/e)(\Omega\cdot \mathrm{cm})^{-1}$  and 357 $(\hbar/e)(\Omega\cdot \mathrm{cm})^{-1}$, respectively. Moreover, we  found that the SHC varied quickly as a function of chemical potential. Because the band crossing points of nodal rings are located near but away from the Fermi level, the curve of SHC $\sigma_{yz}^{x}$ exhibits two peak values below and above the Fermi level, that is, 788 $(\hbar/e)(\Omega\cdot \mathrm{cm})^{-1}$ at $E=E_F-0.035$ eV  and  631 $(\hbar/e)(\Omega\cdot \mathrm{cm})^{-1}$ at $E=E_F+0.075$ eV. These values are approximate to the SHC of Weyl semimetal TaAs $\sim$781 $(\hbar/e)(\Omega\cdot \mathrm{cm})^{-1}$ \cite{PhysRevLett.117.146403}.
Therefore, the V$_6$Sb$_4$ compound is a potential Dirac nodal line semimetal with a large intrinsic SHC, which could provide promising applications for spintronic devices.

\section{summary}
In summary, using first-principles calculations, we theoretically investigated the band topology and the intrinsic spin Hall effect (SHE) of V$_6$Sb$_4$. In absence of the SOC, this compound is a  Dirac nodal line semimetal with symmetry-protected nodal rings near the Fermi level. Within the SOC, the spin-rotation symmetry breaking leads to nodal rings gapped with a small band gap. Furthermore, based on the Wannier tight-binding approach and Kubo formula, we also uncovered the large spin Hall conductivity in the V$_6$Sb$_4$ compound, which is up to 788 $(\hbar/e)(\Omega\cdot \mathrm{cm})^{-1}$ near the Fermi level. The large SHE is attributed to the intrinsic mechanism of spin Berry curvature. Considering that the compound V$_6$Sb$_4$ was recently synthesized in experiments \cite{A166arXiv}, our findings can be expected to facilitate V-based kagome compounds with exotic quantum order for applications in spintronics.

\begin{acknowledgments}
This work is supported by the Anhui Initiative in Quantum Information Technologies (AHY160000), the Key Research Program of Frontier Sciences, CAS, China (QYZDYSSWSLH021)£¬ the National Natural Science Foundation of China (11888101 and 11974062), the National Key Research and Development Program of the Ministry of Science and Technology of China (2017YFA0303001, 2019YFA0704901, and 2016YFA0300201), the Strategic Priority Research Program of the Chinese Academy of Sciences (XDB25000000), and the Science Challenge Project of China (TZ2016004). The DFT calculations in this work are supported by the Supercomputing Center of University of Science and Technology of China.
\end{acknowledgments}

%

\end{document}